\documentstyle[11pt]{article}

\def\lsim{\mathrel{\rlap{\lower4pt\hbox{\hskip1pt$\sim$}}
    \raise1pt\hbox{$<$}}}         
\def\gsim{\mathrel{\rlap{\lower4pt\hbox{\hskip1pt$\sim$}}
    \raise1pt\hbox{$>$}}}         
\newcommand{\be}{\begin{equation}}
\newcommand{\ee}{\end{equation}}

\begin{document}

\setlength{\baselineskip}{0.7 cm}

\begin{center}
{\Large \bf Nuclear Coherent versus Incoherent Effects in Peripheral
 RHI Collisions.}\\[0.3cm]

{\bf M. E. Bracco }\\ [0.1cm]

{\normalsize  Instituto de F\'{\i}sica Te\'orica \\
	      Universidade Estadual Paulista\\
		Rua Pamplona 145  - 01405-900 - S\~ao Paulo, S.P.\\
		Brasil}\\ [0.2cm]
and\\[0.2cm]
{\bf M. C. Nemes} \\[0.1cm]
{\normalsize  Departamento de F\'{\i}sica, ICEX -
		   Universidade Federal de Minas Gerais.\\
		   C.P. 702, Cidade Universit\'aria, 30.000, Belo
		   Horizonte - Minas Gerais\\ Brasil}\\

\end{center}
\vspace{1.0cm}

\begin{abstract}
{\large We derive simple and physically transparent expressions
for the contribution of the strong interaction to one nucleon removal
processes in peripheral relativistic
heavy ion collisions. The coherent contribution,i.e, the excitation
of a giant dipole resonance via meson exchange is shown to be negligible as
well as interference between coulomb and nuclear excitation. Incoherent
nucleon knock out contribution is also derived suggesting the nature of the
nuclear interaction in this class of processes.
We also justify the simple formulae used to fit the
data of the E814 Collaboration.}

\end{abstract}

\section{Introduction}
	The important role played by the electromagnetic interaction in single
nucleon removal processes induced by Relativistic Heavy Ions (RHI) has long
been known.
Dissociation of RHI by the Coulomb field of the target nucleus was first
reported by Heckman and Lindstron \cite{Heckman} using for that purpose
radiochemical techniques. Many other accelerator experiments have followed,
using several combinations of heavy projectiles and targets
with incoming energies $E \gsim 1. $ GeV/nucleon.
The most conspicuous evidence of such electromagnetic
processes is their large cross section and approximate dependence with $Z^2$
, Z being the atomic number. Within this context it is very important to
have a quantitative model which allows one to isolate and identify possible
contributions arising from different processes due, for instance,
to the strong interaction.

 Several simple models have been used to separate
the nuclear contribution the most popular among experimental groups
being the Limiting Fragmentation Model (LFM). This model assumes that the
yield of the particular fragment from the target due to nuclear interactions
will be  independent of the beam, except for a geometrical factor
\cite{Olson}-\cite{Hill3}.
Based on this model a parametrization of the nuclear contribution is
performed and the electromagnetic contribution extracted from the data.
More recently in Brookhaven the same type of process is investigated
\cite{Takai},\cite{Takai1}.
Their data analysis assumes the one nucleon removal to be
basically due to Coulomb dissociation (parametrized as $ b  Z^c_T$, $b$ and
$c$ being free parameters) and to nuclear effects (parametrized as the
sum of the heavy ions' radii ($ a\;(A_T^{1/3} + A_P^{1/3} $),  $ a$
being a free parameter).
In this context an essential question arises: is it possible to give
a sound theoretical basis for the contribution of the strong interaction
in the one nucleon removal process? Can one theoretically justify
the E814 prescription, for the nuclear cross section, quantitatively?
Answering these questions are the main purpose of the present work.

We consider two mechanisms which could be responsible for one nucleon removal
mediated by the nuclear field: the excitation of a giant dipole resonance
via vector meson exchange and the direct incoherent knock-out contribution.
In the first case, i.e., for the collective effects we use the framework
of ref.\cite{Luiz} and derive analytical expressions for pure Coulomb
(rather well known), pure nuclear and interference cross sections
(here obtained for the first time). We show that the theoretical cross
sections involving the strong interaction are independent of the bombarding
energy, depend very
sensitively on the meson mass and display a different qualitative behavior as
compared to the empirically predicted one from E814. Moreover for realistic
values of the coupling constant and mesonic masses these cross sections are
too small. We are therefore led to the conclusion that one nucleon removal
through the excitation of a giant dipole resonance by means of the
nuclear field cannot be the mechanism responsible for the observed nuclear
contribution.

We next investigate the possibility that one nucleon is removed by means
of and incoherent process (fragmentation reaction).
The adequate theoretical tool for this purpose is
Glauber's multiple scattering theory \cite{Glauber}. It has long been
known both for its
simplicity and amazing predictive power. One can find
copious examples in the literature where such theory allows for a simple
physical interpretation of experimental results as well as their
quantitative analysis (\cite{Matthiae}-\cite{caro}).
We therefore work in such scheme. In fact,
 fragmentation reactions of the type discussed here have already been
successfully analyzed in the framework of Glauber's theory: in one
nuclear removal reactions, the momentum distribution of the outgoing
fragment has been shown to reflect the momentum distribution of the
nucleon which is removed from the surface of the projectile nucleus
\cite{caro}. This is precisely the physics behind the incoherent nuclear
one nucleon removal, as we shall show here.
The target dependence so obtained corresponds precisely to
the empirically fitted one.
We are thus able to reproduce the results of the E814 experiment and
 to give a physical interpretation of the empirical fit.
 The calculation presented here, we believe,
 provides for a good theoretical prescription as to the quantitative
 nuclear contribution to the process.

This paper is organized as follows:
in section 2 the Coherent Effects, i.e, the excitation of a Giant Dipole
 Resonance via one photon and one vector meson exchange are analyzed.
Analytical expressions for the pure nuclear and interference contributions
are derived.
Section 3, presents the study of the Nuclear Incoherent Effects,
using Glauber's formalism.
Finally we compare our results with E814 data showing very good agreement.

\vspace{0.3in}
\section{Nuclear Coherent Effects}
\vspace{0.1in}
${ }$

		Recently the E814 Collaboration \cite{Takai}
has reported on the total cross section for the excitations of the
giant dipole resonance by means of peripheral heavy ion collisions.
The reaction considered was $^{28}Si $ into $p + ^{27}Al $ at an energy of $
14.5$  GeV/nucleon  bombarding different targets
($C, Al, Cu, Sn, Pb$).
The experimental data suggest a non negligible amount of nuclear effects
and show that the final-state energy, is peaked near the isovector giant
dipole resonance in $^{28}Si$.

The pure Coulomb contribution is quantitatively the most important
and has been previously obtained by several authors by different
methods (Refs. \cite{Luiz}, \cite{Bertu},\cite{Galleti}). On the other hand
the pure nuclear and interference cross sections are not as
well understood (see ref. \cite{Vary}).

In this section we study the excitation of a giant dipole resonance
via one photon and one meson exchange and assume the one nucleon removal
to be entirely due to this process.

The study of vector and scalar meson exchanges in a covariant
formalism has been performed in detail
in ref. \cite{Airton} for inelastic proton-nucleus scattering.
In this reference it is shown that within the relativistic eikonal
approximation one can sum an infinite series of Feynman diagrams
and the resulting scattering amplitude is strongly
dominated by the vertex where the excitation takes place.
 We therefore restrict ourselves in our derivations to the
lowest order Feynman diagram contributions for the excitation
of a giant dipole resonance via vector meson and photon exchange.

We show that in the ultra-relativistic limit it is possible to derive very
simple analytical formula for the three contributions (pure Coulomb,
pure nuclear and interference), to understand
the physical reason why the pure nuclear contribution is so
much smaller than the Coulomb one and to estimate the contribution
of interference.

\vspace*{0.2in}
\subsection{Collective Coulomb plus Nuclear Excitation Cross Section}
\vspace*{0.1in}

${ }$

Formally, the first order contribution to collective nuclear excitation via
one photon and one meson exchange between projectile and target nucleus is
given in the Center-of-Mass System (C.M.) by \cite{Luiz}:

\begin{eqnarray}
{d\sigma \over d\Omega}\biggr|_{CM}&=&\sum_{f_A} \sum_{f_B}
{M'_A M'_B M_A M_B \over (\sqrt{s})^2}\biggl[ {2 \alpha Z_A Z_B \over q^2} +
{g_v \over q^2 - \mu^2}\biggr]^2 \nonumber \\
& &|F^{\mu}_B (\vec q_B,\lambda_B,m_B) \Lambda^{\mu}_{\nu}
      (P_A \rightarrow P_B) F^{\nu}_A (\vec q_A,\lambda_A,m_A) |^2
\end{eqnarray}
where $\alpha$ is the fine-structure-constant, $\sqrt{s}$ is the total center
of mass energy, the $Z$'s and $M$'s are atomic and mass numbers of each
nucleus,
respectively, $q$ is the four-momentum transfer and $\vec q_A$ ($\vec q_B$)
its spatial part in the proper system of the nucleus $A$ ($B$)
(ref. \cite{Luiz}). $g_v$ is the strong coupling constant and
$\mu$ is the vector meson's mass.

The $F_A^l$ are the Fourier transforms of transition current matrix elements,
given by:

\begin{eqnarray}
F^{l}_{A}(q_A;\lambda_A,m_A)&=&
{1 \over C} \int d^3r < E^*_A;\lambda_A, m_A|
j^{l}_A(\vec {r})|
A g.s > e^{i \vec {q_A} \cdot \vec {r}} \nonumber \\
&=& < E^*_A;\lambda_A, m_A| j^{l}_A(q_A)|A_{g.s}>
\end{eqnarray}
with $C= eZ_A $ for the Coulomb contribution and $C= g_v$ for the strong
contribution. We can write the similar expression for the nucleus $B$.

In the above equations, currents and nuclear states are represented in their
proper systems, and $\vert E;\lambda,m>$ stands for the final state of nucleus
$A$ or $B$ with angular momentum $\lambda$ whose projection in the incident
beam direction $(e'_z$) is $m$. $\sum_f$ means the summation over final
states. The Lorentz transformation matrix is given by
\be
\Delta_{\tau l}=
\left(
\begin{array}{cccc}
\gamma       & 0  & 0   & -\beta \gamma \\
0            & -1 & 0   & 0             \\
0            & 0  & -1  & 0 \\
\beta \gamma & 0  & 0   & -\gamma
\end{array} \right)
\ee

\medskip

\noindent
where $\beta$ is the velocity of nucleus $A$ in the rest frame $B$, and
$\gamma$ is the associated Lorentz factor ($\gamma = \sqrt{1-\beta^2}$).

In order to evaluate the current matrix elements, it is convenient to use the
coordinate system whose z--direction is taken to be the direction of the
momentum transfer seen by the nucleus. The cartesian base vectors
$\{ \hat{e_x},\hat{e_y},\hat{e_z} \}$, where $\hat{z}={q}/|\vec{q}|$,
are related to the original ones $\{ \hat{e'_x},\hat{e'_y},\hat{e'_z} \}$ by

\begin{eqnarray}
\left(
\begin{array}{c}
\vec{e_x} \\
\vec{e_y} \\
\vec{e_z}
\end{array} \right)&=&
\left( \begin{array}{ccc}
(\cos\theta-1)\cos^2\varphi+1 & (\cos\theta-1)\sin\varphi \cos\theta
			 & -\sin\theta \cos\varphi \\
(\cos\theta-1)\sin\varphi \cos\varphi & (\cos\theta-1)\sin^2\varphi + 1
			 & -\sin\theta \sin\varphi \\
\sin\theta \cos\varphi     & \sin\theta \sin\varphi & \cos\theta
\end{array} \right) \nonumber\\
& & \times
\left( \begin{array}{c}
\vec{e'_x} \\
\vec{e'_y} \\
\vec{e'_z}
\end{array} \right)
\end{eqnarray}

\medskip

\noindent
where $\theta$ and $\varphi$ are polar angles of $\hat{z}$ with respect to the
base vectors $\{ \hat{x'},\hat{y'},\hat{z'} \}$. The corresponding spherical
base vectors $\{ \hat{e_{-1}},\hat{e_0},\hat{e_{+1}} \}$ and
$\{ \hat{e'_{+1}},\hat{e'_0},\hat{e'_{-1}} \}$ are related to each other by

\be
\left(
\begin{array}{c}
\hat{e}_{+1} \\
\hat{e}_0 \\
\hat{e}_{-1}
\end{array} \right) =
\left( \begin{array}{c}
\hat{e}'_{+1} \\
\hat{e'}_0 \\
\hat{e'}_{-1}
\end{array} \right)
D^{(1)}(\varphi,\theta,-\varphi)
\ee
where $D^{(1)}(\varphi,\theta,-\varphi) $ is the usual rotation
matrix for total angular momentum
$j=1$. Using these relations we can express the four-vector current in terms
of spherical components as

\be
\left( \begin{array}{c}
\rho \\
j'_x \\
j'_y \\
j'_z
\end{array} \right) =
\left( \begin{array}{cccc}
1 & 0 & 0 & 0 \\
0 &   &   &   \\
0 &   & UD^T & \\
0 &   &   &
\end{array} \right)
\left( \begin{array}{c}
\rho \\
j_{+1} \\
j_0 \\
j_{-1}
\end{array} \right)
\ee
where the $3 \times 3$ matrix $U$ transforms spherical vectors into cartesian
ones. The convenience of the spherical basis now becomes apparent: the
selection rules for the current matrix elements are

$$ <E^*;\lambda,\nu|\rho|0,0> = <\rho> \delta_{\nu,0}, $$
$$ <E^*;\lambda,\nu|j_{+1}|0,0> = <j_{+1}> \delta_{\nu,+1}, $$
\be <E^*;\lambda,\nu|j_{0}|0,0> = <j_{0}> \delta_{\nu,0}, \ee
$$ <E^*;\lambda,\nu|j_{-1}|0,0> = <j_{-1}> \delta_{\nu,-1}
				= <j_{+1}> \delta_{\nu,-1},$$
where $\nu$ refers to the projection of angular momentum in the three
$\vec{q}$ directions.

We can now write the matrix element in eq. (1) as

\begin{eqnarray}
M&=&\bar{D}^{(\lambda_B)} (\hat{q}_B)
\left( \begin{array}{cccc}
    0    & <j^B_{+1}> &    0    &      0     \\
<\rho^B> &      0     & <j^B_0> &      0     \\
    0    &      0     &    0    & <j^B_{-1}>
\end{array} \right) \times  \nonumber \\
& & \left( \begin{array}{cccc}
1 & 0 & 0 & 0 \\
0 &   &   &   \\
0 &   & D^{(1)\dagger}(\hat{q}_B) &   \\
0 &   &   &
\end{array} \right) \times
\left( \begin{array}{cccc}
\gamma        & 0 & -\beta{\gamma} & 0 \\
0             & 0 & 0       & 1 \\
\beta{\gamma} & 0 & -\gamma & 0 \\
0 & 1 & 0 & 0
\end{array} \right) \times
\left( \begin{array}{cccc}
1 & 0 & 0 & 0 \\
0 &   &   &   \\
0 &   & D^{(1)^*}(\hat{q}_A) &   \\
0 &   &   &
\end{array} \right)\nonumber\\
& & \times
\left( \begin{array}{ccc}
    0      &  <\rho^A> &    0      \\
<j_{+1}^A> &      0    &    0      \\
    0      & <j^A_{0}> &    0      \\
    0      &      0    & <j^A_{-1}>
\end{array} \right) \bar{D}^{(\lambda_A)T}(\hat{q}_A)
\end{eqnarray}

\medskip

\noindent
$ \hat{q}$ (the argument of $ D(\hat q) $), symbolically represents the
Euler angles associated to the rotation required to transform $\hat{e'_z}$
to $\hat{e_z}$ in each nucleus.

In the case where the nucleus $B$ stays unexcited, we obtain

\begin{eqnarray}
{d\sigma \over d \Omega}\biggr|^{single}_{m=\pm 1}&=&{1 \over 2} F_c
|<\rho_B>| ^2 \gamma^2 \nonumber \\
& &\sin^2 \theta_A
|[<\rho_A>  - \beta \cos{\theta_A} <j_0^A>]
+ \beta \cos{\theta_A} <j^A_1>|^2
\end{eqnarray}

\begin{eqnarray}
{d\sigma \over d \Omega} \biggr|^{single}_{m=0}&=&F_c |
<\rho_B>| ^2 \gamma^2 \nonumber \\
& & | \cos{\theta_A} \{<\rho_A> - \beta \cos{\theta_A} <j_0^A>\}
-{1 \over 2} \beta \sin^2{\theta_A} <j^A_1>|^2
\end{eqnarray}
where $\theta_A$ represents the polar angle of $\vec{q_A}$ in the
rest frame of $A$ and similar expressions can be constructed for the strong
and interference part with the respective substitution of $F_c$ by $F_s$
and $F_{cs}$.

The next problem one has to face in the present formulation is the inclusion
of nuclear absorption effects. The way to circumvent this problem is the
following: we first write the total cross-section in terms of an integral over
the transverse momentum transfer,
\be
\sigma_{tot}= \int d^2 q_{\perp}
\biggl[{\sqrt{F_c}\over q^2}+{\sqrt{F_s} \over
q^2 - \mu^2}\biggr]^2 Tr[ M^{\dagger}(\vec q_{\perp}) M(\vec q_{\perp})]
\ee
where $\sqrt{F_c}=2 \alpha Z_A Z_B $ and $\sqrt{F_s}=g_v $.
We then introduce a Fourier transform
\be
f(b) = {1 \over 2\pi} \int d^2 \vec q_{\perp}
\biggl[{\sqrt{F_c}\over q^2}+{\sqrt{F_f} \over
q^2 - \mu^2}\biggr] \sqrt{M(\vec q_{\perp})} e^{i\vec q_{\perp} \cdot \vec b}
\ee
where $M(\vec q_\perp)$ denotes

\begin{eqnarray}
M(\vec q_{\perp})&=&\sqrt{{d^2\Omega \over dq_{\perp}^2}} M(\vec q)
\nonumber\\
&=& {1 \over \sqrt{p\biggl(\sqrt{p^2-q^2_{\perp}}
\biggr)}}M(\vec q)
\end{eqnarray}

\noindent and $p$ stands for the projectile's center-of-mass momentum.
The total cross section can now be written as

\be
\sigma_{tot}=\int d^2 b Tr[f^{\dagger}(b) f(b)]
\ee

\medskip

\noindent
The above formula can be interpreted as the impact parameter
representation of the cross section. It is now possible to estimate nuclear
absorption effects in a simple way: we attribute those parts of the
integral coming from all values of $b$ smaller than the sum of radii
of projectile and target to the absorption coming from nuclear
interactions which do not contribute to excite the collective mode.
Thus, with $b_{min}= r_0 (A_T^{1/3} + A_P^{1/3})$,
we get:

\begin{eqnarray}
\sigma_{tot}&=&{1 \over (2\pi)^4} \int_{b_{min}}^{\infty} d^2 b
\int d^2 q_{\perp} \int d^2 q'_{\perp}
\biggl[{F_c \over q^2(q_{\perp}) q^2(q'_{\perp})} + \nonumber \\
& & 2 {\sqrt{F_c F_s} \over q^2(q_{\perp}) (q^2(q'_{\perp})-\mu^2)}
+ {F_s \over  (q^2(q_{\perp})-\mu^2)(q^2(q'_{\perp})-\mu^2)}\biggr]
\nonumber\\
& & Tr\biggl[M^{\dagger}(\vec q_{\perp})
e^{-i(\vec q_{\perp}-\vec q'_{\perp})\cdot \vec b}
M(\vec q'_{\perp})\biggr]
\end{eqnarray}
where

\begin{eqnarray}
\sigma_{c}&=&{1 \over (2\pi)^4} \int_{b_{min}}^{\infty} d^2 b
\int d^2 q_{\perp} \int d^2 q'_{\perp}
\biggl[{F_c \over q^2(q_{\perp}) q^2(q'_{\perp})}\biggr] \nonumber\\
& & Tr\biggl[M^{\dagger}(\vec q_{\perp})
e^{-i(\vec q_{\perp}-\vec q'_{\perp})\cdot \vec b}
M(\vec q'_{\perp})\biggr],
\end{eqnarray}

\begin{eqnarray}
\sigma_{s}&=&{1 \over (2\pi)^4} \int_{b_{min}}^{\infty} d^2 b
\int d^2 q_{\perp} \int d^2 q'_{\perp}
\biggl[
 {F_s \over  (q^2(q_{\perp})-\mu^2)(q^2(q'_{\perp})-\mu^2)}\biggr]
\nonumber\\
& & Tr\biggl[M^{\dagger}(\vec q_{\perp})
e^{-i(\vec q_{\perp}-\vec q'_{\perp})\cdot \vec b}
M(\vec q'_{\perp})\biggr]
\end{eqnarray}
and

\begin{eqnarray}
\sigma_{cs}&=&{1 \over (2\pi)^4} \int_{b_{min}}^{\infty} d^2 b
\int d^2 q_{\perp} \int d^2 q'_{\perp}
\biggl[
 2 {\sqrt{F_c F_s} \over q^2(q_{\perp}) (q^2(q'_{\perp})-\mu^2)}\biggr]
\nonumber\\
& & Tr\biggl[M^{\dagger}(\vec q_{\perp})
e^{-i(\vec q_{\perp}-\vec q'_{\perp})\cdot \vec b}
M(\vec q'_{\perp})\biggr],
\end{eqnarray}

\medskip

correspond to the Coulomb excitation process,
to the pure strong contribution and to the
interference between two, respectively.

\vspace*{0.2in}
\subsection{Results}
\vspace*{0.1in}

${ }$

In order to obtain analytical results we now need a model for the
matrix elements corresponding to the excitation of a giant dipole
resonance. We use a macroscopic model. T. Suzuki and D. J. Rowe \cite{Suzuki}
have derive a sum rule for current matrix elements which corresponds
to the incompressible fluid model. In it, the giant
resonance state is represented by a unique level of given multipolarity
and the transition matrix elements of the current from the ground state
to this level is assumed to be of the form

\be
<g.s.|J(\vec r)|E^*, \lambda \nu> = {N \over {2im}} \rho_0(\vec r)
\bigtriangledown \{\vec r^{\lambda} Y_{\lambda \nu}\}
\ee
where $\rho_0( \vec r)$ is the ground state density distribution,
$m$ the nucleon mass and $N$ the normalization constant related to the
sum rule value $S$ as,

\be
N=\sqrt{E^*/S}
\ee
with
\be
S = 3 \lambda A R^{2 \lambda -2} / 8 \pi m
\ee
The Fourier transform of eq. (19) gives the required matrix element. After
some algebra we obtain

\be
<\rho> = {N \sqrt{4 \pi (2 \lambda +1)} \lambda \int dr r^{\lambda+1}
\rho_0 (r) j_{\lambda-1}(qr) \over 2 m i^{\lambda} E^{*} / |\vec{q}| }
\ee
where $j_{\lambda+1}(qr)$ are spherical Bessel functions and

\be
<j_0>={E^* \over |q|} <\rho>,
\ee

\be
<j_1>=\biggl[{\lambda +1 \over 2\lambda}\biggr]^{1/2} <j_0>
\ee

In this model, magnetic current is neglected. It is known that for giant
resonances, the contribution of magnetic current is small. For very high
energies ($\gamma >> 1$) and in the low momentum transfer limit
($qR << 1$), we may approximate the nuclear matrix element in eq. (22) as

\be
<\rho> = \sqrt{ {A E^{*} \over 2m}} {|q_A| \over E^{*} }
\ee
Furthermore, the four-momentum transfer squared is given by

\be
q^2 \simeq - \biggl[ \biggl( {E^{*} \over \beta \gamma} \biggr)^2
	   + q^2_{\perp} \biggr]
\ee
In this limit,

\be
M_0 (q_{\perp}) \sim 0
\ee
and
\be
M_1 (q_{\perp}) \sim {\gamma \over p} (A E^{*} / 2m)^{1/2} q_{\perp} e^{-iq}
\ee
and it is possible to carry out the integrals in $q_{\perp}$ and
$q'_{\perp}$
as well as over the impact parameter $b$. We obtain for the Coulomb
contribution

\be
\sigma_c = 3.5 \times 10^{-3} Z^2_B {N_A Z_A\over A_A^{2/3} }
	   \ln{[\delta / \xi +1]} \; mb.
\ee
\medskip
\noindent
where $\xi = E^{*} b_{min} / \beta \gamma$ and $\delta = 0.681085 \ldots$.

The pure strong interaction and interference contributions can be obtained in
an analogous manner. We get

\be
\sigma_s = 26.65  (2 \pi^2) g^4_v A_A^{4/3} e^{-2 \mu b_{min}}\; mb
\ee
and
\be
\sigma_{cs} = {\alpha g_v^2 (\pi N_A Z_A)^{1/2} \over 10 m} Z_B A_B^{1/3}
	     { e^{-\mu b_{min}} \over \sqrt{\mu b_{min}} }  \;mb
\ee

\medskip

\noindent
As discussed before, expression (29) has been obtained before by several
authors by different methods.
The results for the strong ($\sigma_s$) and interferency part ($\sigma_{cs}$)
, eqs. (30) and (31), we believe to be novel.

{}From the qualitative point of view it is clear that the theoretical
prediction for the $b_{min} = r_0 (A_A^{1 \over 3} +A_B^{1 \over 3})$
dependence of the total strong cross section (eq.(30)) is in contradiction
with the empirical parametrizations obtained in the E814 collaboration given by
(Ref.\cite{Takai}):

\be
\sigma_{exp-s} = a (A_A^{1/3}+A_B^{1/3})
\ee
 where $a = 1.34 \pm 0.19 fm^2 $.
The theoretical cross sections depend very sensitively on the realistic
values of the exchanged meson mass $\mu (MeV) $ and of the coupling
constant ${g_v^2 \over (4\pi)} = 0.9 $ \cite{Ollinde}. For these reasons
the nuclear and interference contributions to the collective excitation
become negligible. Also as expected these two contributions are energy
independent. The conclusion then is that the collective effects are essentially
due to coulomb excitation and that the proposed mechanism is in
rather good agreement with the data (see table 1, columns 2 and 4).
Notice, however, that the experimental values for the strong part of the
interactions are non negligible. This indicates that one should rather
investigate incoherent contributions to the nuclear part. In fact it can
be shown that the incoherent knock-out process gives both qualitative and
quantitative agreement with the empirical findings.

\begin{table}
\begin{center}
\begin{tabular}{|c|c|c|c|c|c|}
\hline
\multicolumn{1}{|c|}{target}&
\multicolumn{1}{c|}{$\sigma_{exp-c}$}&
\multicolumn{1}{c|}{$\sigma_{exp-s}$}&
\multicolumn{1}{c|}{$\sigma_{c}$}&
\multicolumn{1}{c|}{$\sigma_{cs}$}&
\multicolumn{1}{c|}{$\sigma_{s}$}\\
\hline
C  &$ 6.81  \pm 1.03$  &$ 7.13 \pm 1.00$   & 6.34  & $ 1.13 \times 10^{-2}$
&$7.00 \times 10^{-3}$ \\
Al &$ 28.15 \pm 4.88 $ &$ 8.08 \pm 1.14$  & 30.71  & $ 6.68 \times 10^{-3}$  &
$1.70 \times 10^{-3} $\\
Cu &$ 122.79\pm 24.16$ &$ 9.40 \pm 1.33$   & 131.04& $ 2.63 \times 10^{-3}$  &
$1.69\times 10^{-4}$\\
Sn &$ 333.87\pm 70.92$ &$ 10.64 \pm 1.50$  & 366.79& $ 9.14 \times 10^{-4}$
&$1.51 \times 10^{-5}$\\
Pb &$ 828.25\pm 187.65$&$ 12.00 \pm 1.70$  & 927.11& $ 2.50 \times 10^{-4}$
&$9.01\times 10^{-7}$\\
\hline
\end{tabular}
\caption{ {\it Experimental Coulomb ($\sigma_{exp-c}$),
Experimental Strong ($\sigma_{exp-s}$), our Coulomb
($\sigma_{c}$, eq.29) ,Interference ($\sigma_{cs}$, eq.31),
and our Strong cross section  in the last column ($\sigma_{s}$, eq.30),
with $g_v =3.36 $ and $\mu = 300 MeV $.Units in mb.}}
\end{center}
\end{table}

\vspace*{0.2in}
\subsection{Conclusions for Coherent Effects}
\vspace*{0.1in}

${ }$

In this section we use a covariant formulation to describe the
excitation of a collective mode via photon and meson exchange.
Besides the well known total Coulomb contribution, analytical
results are obtained for the total strong interaction contribution and the
interference cross section.
 We show that the fact that the strong interaction is
 mediated by massive mesons drastically reduces its contribution to the
excitation of the giant resonance. Both the pure strong and interference
contributions are independent of the incident energy.

Moreover the theoretical prediction for the strong part of the
collective total cross section is in qualitative disagreement with
the experimental findings.

	Therefore we conclude, that the electromagnetic effect is
	the only responsible for the processes where coherent effects
	are produced, as a giant dipole resonance.
The strong interaction contribution is
coming from incoherent effects such as one nucleon knock-out.
This point is investigated in the next section.

\vspace*{0.2in}
\section{Nuclear Incoherent Effects}
\vspace*{0.1in}

${ }$

In this section we analyse Glauber's formulation of the one nucleon removal
process according to reference \cite{caro}, and give a theoretical
expression for the total nuclear cross section.
We show its compatibility with the empirical
formula used in the analysis of the E814 experiment \cite{Takai}.

	The physical process considered is a nucleon knock-out off the
projectile within the context of Glauber's formalism.

\vspace*{0.2in}
\subsection{The Nuclear One Nucleon Removal Cross Section}
\vspace*{0.1in}

${ }$

In this section the contribution of the strong interaction for the one
nucleon removal process is calculated using Glauber's multiple scattering
formalism \cite{caro}. The notation is summarized in the following equation,
valid
in the projectile's rest frame:

\begin{eqnarray}
^{A}Z\{\vert \vec 0,\psi_0 >\} + T\{\vert -\vec p_0,\theta_0>\}
&& \rightarrow ^{A-1}Z\{\vert \vec k,\phi_{\alpha}>\} \nonumber\\
&& + n \{\vert \vec p,\eta_{\vec p}>\} + T'
\{\vert -\vec p_0 -\vec q,\theta_{\beta}\vert >\}
\end{eqnarray}
Before the reaction, the projectile $^{A}Z$, with intrinsic wave
function $\psi_0$ is at rest and the target approaches with momentum
$-\vec{p_0}$. After the interaction, a fragment with $A-1$ nucleons is
detected in a particle stable state $\phi_{\alpha}$ with momentum
$\vec{k}$. The nucleon with momentum $\vec p$ and scattering wave function
$\eta_{\vec{p}}$, and the final state $\Theta_{\beta}$ of the target,
are unobserved. The cross-section corresponding to the equation (33)
is given by \cite{fujita}

\begin{eqnarray}
{d^3\sigma \over dk^3}&=&\int d^2q \sum_{\alpha}\sum_{\beta}\nonumber\\
& & \vert \int {d^2b \over 2 \pi }
e^{-i \vec q \cdot \vec b}
\biggl< \eta_{\vec q - \vec k};\Theta_{\beta}
\vert 1- \prod_{i \in P, j \in T} (1 - \Gamma_{ij})\vert
\psi_0;\Theta_0 \biggr> \vert ^2
\end{eqnarray}
In the above formula $ \Gamma_{ij}(\vec x_i + b - \vec y_j) $ are
the profile functions for collisions
between a projectile nucleon $i$ located at $\vec{x_i}$ and a target
nucleon $j$ at $\vec{y_j}$; $\vec{x_i}$ and $\vec{y_j}$ refer to the
respective centers of mass and the impact parameter $\vec{b}$ is the
transverse distance between projectile and target.

The fragmentation cross-section corresponding to the incoherent process
is derived in detail in reference \cite{caro}. We quote the result

\be
{d^3\sigma \over dk^3}=\int D(\vec s) d^2 s \int dz \int d^3k W(\vec s,z,\vec
k)
\ee
where $\vec s$ is the perpendicular distance of each nucleon to the center
of the nuclei.
The distorting function $D(\vec{s})$ contains the reaction dynamics

\be
D(\vec s)=\int d^2b e^{-2 Im\chi_{FT}(\vec b)-2 Im \chi_{T}(\vec b+\vec s)}
(e^{\sigma_{NN}(0) T_T(\vec b+\vec s)} -1)
\ee
where
\be
i\chi_{FT}(\vec b) \approx -\langle \phi_{\alpha}\Theta_0 \vert \sum_{m \geq 1}
\Gamma_{mn} \vert \phi_{\alpha}\Theta_0 \rangle
\ee

\be
i\chi_{T}(\vec b + \vec s) \approx -\langle \Theta_0 \vert \sum_{n}
\Gamma_{1n}(\vec b + \vec s - \vec t_n) \vert \Theta_0 \rangle
\ee
\be
T_{T}(\vec B)= A_T \int dz \rho(\vec B,z)
\ee
with $\rho(\vec B,z)$ is the nucleon density function,
and
\be
\sigma_{NN} = \int d^2 q e^{i\vec q \cdot \vec b} {d\sigma_{NN} \over
d^2q}
\ee
is the nucleon-nucleon cross section.

As discussed in reference (\cite{caro}) shown below, $D(\vec{s})$ localizes the
reaction to the nuclear surface. The Wigner transform $ W(\vec s, z, \vec k)$
contains the
momentum distribution of the removed nucleon.

In the calculations we have used Gaussian densities,
\be
\rho(\vec r) = N_g e^{-(r^2/a_0^2)}
\ee
where
for light nuclei we have adjusted $ a_0$ in order to reproduce the root
mean square radius. For heavier nuclei ($ A > 40 $) $N$ and $a_0$ have been
adjusted in order to reproduce the tail of the Woods-Saxon density.

The theoretical results to be shown are based on the result

\be
\sigma_N= \int 2 \pi s D(\vec s) T_p(\vec s) ds
\ee
with
\be
T_p(\vec s)=\int dz \int_{k_{min}}^{\infty} d\vec k \; W(\vec s, z, \vec k)
\ee
$k_{min}$ corresponds to the Coulomb barrier cut-off parameter, in the
case where the nucleon removed is charged ($\sigma_N $ is the nuclear cross
section).

\vspace*{0.1in}
\subsection{Results - E814 Brookhaven Experiment}
\vspace*{0.05in}

${ }$

In the E814 Experiment was reported the results of the EMD of $^{28}Si$ into
$p + ^{27}Al $, and  $ Pb, Sn, Cu, Al$ and $C$ were used as targets.
The total cross section for producing $p + ^{27} Al$ was measured.
Presenting the first measurements of the final-state energy spectrum (for all
targets)
which is peaked near the isovector giant-dipole resonance in $^{28}Si$. The
incident energy was $14.5$ GeV/nucleon.
 A detailed description of the experiment can be found in ref.\cite{Takai}.
The empirical fitting was for the electromagnetic part:

\be
\sigma_{exp-c} = b Z_T^c
\ee
and for the nuclear part:
\be
\sigma_{exp-s} = a (A_T ^{1/3} + A_P^{1/3})
\ee
with $a=1.34\pm 0.19 fm^2., b=0.23\pm 0.05 fm^2.$ and $c=1.8\pm 0.06.$.
Using this fitting the authors showing a good agreement for all ions.
We calculate the nuclear incoherent cross section (eq.42), using the
corresponding cut-off for the Coulomb barrier.

	Before a comparison between the theoretical results and
the E814 data a word about the adequacy of such theoretical description
to this specific set of data is in order: in the E814 experiment events
which correspond to an energy transfer above 200 MeV have been excluded
precisely in order to avoid contributions coming from the nuclear
interaction. In Glauber's model the energy transfer is realized by
means of nucleon-nucleon collisions which are assumed to occur
essentially in the impact parameter plane. The average momentum
transfer is 400 $MeV/c$ which corresponds to and energy of about 80 $MeV$.
Also the highest probabilities occur for smaller momentum transfers.
We can therefore not exclude such processes from the experiment.
The qualitative and quantitative agreement with experiment is,
in the end what strengthened our conclusion.

	In Table 2 we can see the nuclear cross section fitted by the experimentalists
and our theoretical calculation of the nuclear
part. We find good agreement. In this way our conclusion is that the processes
is due to incoherent effects. Our numerical result based in eq. (42) gives
a nuclear cross section proportional to the sum of the radii of the
target and projectile, this is showing in the third coulumn of this table
where the ratio $ {\sigma_N \over (A_P^{1/3}+A_T^{1/3})}$ is shown to remain
constant.

	We demonstrated that in this class of process, the interference
cross section between Coulomb and strong interaction is negligible.
The nuclear part is due to an incoherent process. In the context of
Glauber's theory, we show that the nuclear cross section is proportional
to the sum of the radii of nuclei, in perfect agreement with the experimental
fitting used in E814 Collaboration.

	In table 3 we can see the total cross section obtained in the
E814 Collaboration in comparison with our theoretical results for the
electromagnetic coherent and nuclear incoherent parts. Good agreement
is found.

\begin{table}
\begin{center}
\begin{tabular}{|c|c|c|c|}
\hline
\multicolumn{1}{|c|}{$A_T$}&
\multicolumn{1}{c|}{$\sigma_{exp-s}(E814)$}&
\multicolumn{1}{c|}{$\sigma_N$} &
\multicolumn{1}{c|}{ ${\sigma_N} / ({A_P^{1/3} + A_T^{1/3}) } $}\\
\hline
$^{12}C$  & $7.11 \pm 1.00$ & 8.11 & 1.5\\
$^{27}Al$ & $8.04 \pm 1.14$ & 9.24 &1.5 \\
$^{63}Cu$ & $9.34 \pm 1.33$ & 11.12 & 1.5\\
$^{118}Sn$& $10.59\pm 1.50$& 13.04 & 1.5 \\
$^{208}Pb$& $11.95\pm 1.70$  & 15.23 & 1.6 \\
\hline
\end{tabular}
\caption{{\it First column shows the parametrization used in E814
(nuclear part),
second column our results, third column shows the proportionality with the
sum of the radii.}}
\end{center}
\end{table}

\begin{table}
\begin{center}
\begin{tabular}{|c|c|c|c|}
\hline
\multicolumn{1}{|c|}{$A_T$}&
\multicolumn{1}{c|}{$\sigma_{exp-t}(E814)$}&
\multicolumn{1}{c|}{$\sigma_N+ \sigma_{cs}$} \\
\hline
$^{12}C$  & $13.92  \pm 2.03$ & 14.45  \\
$^{27}Al$ & $36.19  \pm 6.02$ & 39.95   \\
$^{63}Cu$ & $132.13 \pm 25.49$ & 142.16  \\
$^{118}Sn$& $344.46 \pm 72.47$ & 349.83  \\
$^{208}Pb$& $840.20 \pm 189.35$ & 942.34  \\
\hline
\end{tabular}
\caption{{\it First column shows the total cross section data of the E814
Collaboration,
second column show the sum of our results for the nuclear and electromagnetic
cross section.}}
\end{center}
\end{table}

\vspace*{0.2in}
\section{Conclusions}
\vspace*{0.1in}

${ }$

In the present work the nuclear and interference nuclear-coulomb contribution
to one nucleon removal processes
in peripheral RHI collisions are studied. Firstly its coherent contribution is
analyzed assuming
that a giant dipole resonance is excited via one vector meson exchange. To
lowest order simple analytical expression for the electromagnetic, nuclear and
coulomb-nuclear interference contributions are derived.

For the electromagnetic part a cross section proportional to
$\approx Z^2$ is obtained, as is expected experimentally and theoretically.
The interference and coherent nuclear part are small and in qualitative
disagreement
with the data. We therefore conclude that the nuclear contribution in
experiment of one nucleon
dissociation, results from incoherent nucleon-nucleon collisions.
Therefore, in section 3,  the incoherent one nucleon knock-out is
studied within the context of Glauber's formalism.
The total nuclear contribution so obtained is shown to be proportional
to the sum of the radii of the nuclei as could be intuitively expected.
Quantitatively speaking the incoherent process is shown to be of the
order expected in experiments.

\section{Acknowledgements}

${ }$

The work of MEB and MCN was partially supported by the Brazilian agencies
FAPESP and CNPq.
	The authors wish to express their gratitude to Prof. Dr. T. Kodama
for valuable discussions throughout the realization of the present work.

\pagenumbering{arabic}

\end{document}